\documentclass[useAMS,usenatbib,fleqn]{mn2e}
\usepackage{graphicx}
\usepackage{epsfig}
\usepackage{color}
\usepackage{tabularx}
\usepackage{multirow}
\usepackage[para,online,flushleft]{threeparttable}
\usepackage[ampersand]{easylist}
\usepackage{amsmath}
\usepackage{hyperref}
\usepackage{ulem}
\hypersetup{
    colorlinks=true,
    linkcolor=blue,
    filecolor=blue,      
    urlcolor=blue,
    citecolor=blue,
    filecolor=blue,
}
\usepackage{graphicx}	
\usepackage{amsmath}	
\usepackage{amssymb}	

\def\psrj1906{PSR J1906 + 722}
\def\psrj1907{PS J1907.3 + 706}

\date{Accepted 2020 December 7. Received 2020 November 12; in original form 2020 August 7.}
\pubyear{2020}

\title[Star formation origin of gamma rays from $\it{3FHL~J1907.0+0713}$]
{Probing the star formation origin of gamma rays from 3FHL J1907.0+0713}

\author[T.Ergin et al.]{T.~Ergin$^{1}$\thanks{E-mail: tulun.ergin@tubitak.gov.tr (TE); labsaha@ucm.es (LS); pooja.bhattacharjee@jcbose.ac.in (PB); hidetoshi.sano@nao.ac.jp (HS); sjtanaka@phys.aoyama.ac.jp (ST)},
L.~Saha$^{2}$, P.~Bhattacharjee$^{3,4}$, H.~Sano$^{5}$, S.~J.~Tanaka$^{6}$, P.~Majumdar$^{8}$, \newauthor
R.~Yamazaki$^{6,7}$, Y.~Fukui$^{9}$
\\
$^{1}$ TUBITAK Space Technologies Research Institute, ODTU Campus, 06800, Ankara, Turkey\\
$^{2}$ IPARCOS and Department of EMFTEL, Universidad Complutense de Madrid, E-28040 Madrid, Spain\\
$^{3}$ Centre for Astroparticle Physics and Space Science, Bose Institute, Block EN, Sector V, Salt Lake, Kolkata 700091, India\\
$^{4}$ Department of Physics, Bose Institute, 93/1 A.P.C. Road, Kolkata 700009, India\\
$^{5}$ National Astronomical Observatory of Japan, Mitaka, Tokyo 181-8588, Japan\\
$^{6}$ Department of Physics and Mathematics, Aoyama Gakuin University, 5-10-1 Fuchinobe, Sagamihara 252-5258, Japan\\
$^{7}$ Institute of Laser Engineering, Osaka University, 2-6 Yamadaoka, Suita, 565-0871 Osaka, Japan\\
$^{8}$ Saha Institute of Nuclear Physics, HBNI, 1/AF Bidhannagar, Kolkata 700064, India\\
$^{9}$ Department of Physics, Nagoya University, Chikusa-ku, Nagoya, Aichi 464-8601, Japan 
}

\begin{document}
\date{}
\pagerange{\pageref{firstpage}--\pageref{lastpage}} \pubyear{2020}
\maketitle
\label{firstpage}

\begin{abstract}
Star-forming (SF) regions embedded inside giant molecular clouds (GMCs) are potential contributors to Galactic gamma rays. The gamma-ray source 3FHL J1907.0+0713 is detected with a significance of roughly 13$\sigma$ in the 0.2 $-$ 300 GeV energy range after the removal of gamma-ray pulsation periods of PSR J1906+0722 from the {\it Fermi}-LAT data set of about 10 years. The energy spectrum of 3FHL J1907.0+0713 is best-fitted to a power law model with a spectral index of 2.26 $\pm$ 0.05. The CO($J$ = 1$-$0) data taken by NANTEN2 revealed that 3FHL J1907.0+0713 is overlapping with a GMC having a peak velocity of about 38 km s$^{-1}$. The best-fitting location of 3FHL J1907.0+0713 is measured to be approximately 0.13 degrees away from the Galactic supernova remnant (SNR) 3C 397 and it overlaps with a star that is associated with a bow-shock nebula. We show that there is no physical connection between 3FHL J1907.0+0713, 3C 397, as well as any positional coincidence with the pulsar. The spectrum of 3FHL J1907.0+0713 is fitting to both hadronic and leptonic gamma-ray emission models and the total luminosity at a distance of 2.6 kpc is calculated to be 1.1 $\times$ 10$^{34}$ erg s$^{-1}$. We also discuss possible SF origins of gamma rays from 3FHL J1907.0+0713, where SNRs, massive protostar outflows, stellar winds from runaway stars, colliding wind binaries, and young stellar clusters are considered as candidate sources.
\end{abstract}

\begin{keywords}
ISM: individual objects: 3FHL J1907.0+0713 (4FGL J1906.9+0712) -- pulsars: individual: PSR J1906+0722 -- ISM: individual objects: 3C 397 (G41.1-0.3) -- stars: winds, outflows -- gamma-rays: ISM -- submillimetre: ISM.
\end{keywords}


\section{Introduction}
\label{Section1}
Giant molecular clouds (GMCs) and dense cloud cores hold the parent material from which stars are formed. These star-forming regions (SFRs) contain newly forming stars (YSOs), where a single or several of such YSOs can illuminate the massive dark cloud in high-energy gamma rays through the interactions of their proto-stellar accretion disks and collimated thermal jets \citep{Ro10}. Typically, strong IR luminosities, non-thermal radio emission and maser activity are the best tracers for YSOs. Once a SFR ages, early-type hot and massive stars are formed that include Wolf-Rayet (WR) and OB stars with strong stellar winds \citep{Ah19}. These newly formed massive stars are usually leading to the formation of H\,{\sc ii} regions inside GMCs. Cosmic rays (CRs) are thought to be accelerated in single, binary, or collective processes of these massive star winds, which result in gamma-ray production \citep{Re06}. A colliding wind binary system, e.g. WR 140 \citep{PiDo06}, may be a significant gamma-ray source with a luminosity up to 10$^{34}$ erg s$^{-1}$ at E $>$ 100 MeV. The best indication of efficient particle acceleration in such objects is the clear detection of non-thermal radio emission from the colliding wind region. Protons might escape the system illuminating nearby MCs. The collective activities of massive stars may form super-bubbles, where the acceleration of multiple shocks can raise the maximum energy of CR protons above one PeV \citep{Ah19}. Other rather well-studied gamma-ray production sites related to SFRs are pulsars and their synchrotron nebulae and supernova explosions and their expanding remnants \citep{Be07}. So far, the {\it Large Area Telescope} on board {\it Fermi Gamma Ray Space Telescope} ({\it Fermi}-LAT) has detected gamma rays from various SFRs in the Galaxy, such as Cygnus OB2 or the Cygnus Cocoon \citep{Ac11}, NGC 3603 \citep{YaAh17, Sa20}, Westerlund 2 \citep{Ya18}, W43 \citep{Le08, YaWa20}, and W40 \citep{Su20}, some of which were also detected by the very high energy gamma-ray observatories(e.g. Cygnus Cocoon: \citet{Ho17}, Westerlund 2: \citet{Ah17}).

3FHL J1907.0+0713 is a point-like gamma-ray source listed in the 3rd hard sources (3FHL) catalog \citep{Aj17} of {\it Fermi}-LAT. It was revealed during the off-pulse analysis of the young and energetic isolated pulsar PSR J1906+0722, which was discovered by Einstein@Home\footnote{\url{https://einsteinathome.org/}} during the blind search study of unidentified gamma-ray sources detected by {\it Fermi}-LAT \citep{Cl15}. PSR J1906+0722 was found to be associated with 3FGL J1906.6+0720, one of the brightest gamma-ray sources in the 3rd {\it Fermi}-LAT point sources (3FGL) catalog \citep{Ac15} and was also listed in the 4th {\it Fermi}-LAT point sources (4FGL) catalog as 4FGL J1906.4+0723. 3FHL J1907.0+0713 was listed as 4FGL J1906.9+0712 in the 4FGL catalog.

Initially, \citet{Cl15} suggested that the excess of gamma-ray emission, whose location overlaps with the position of 3FHL J1907.0+0713/4FGL J1906.9+0712, was about 0.$\!\!^{\circ}$23 away from PSR J1906+0722/4FGL J1906.4+0723 and was possibly not related to this pulsar. Instead, they suggested that it may be a result of the interaction between the SNR 3C 397 (G41.1-0.3) and molecular clouds (MCs) due to its closeness to 3C 397 (about 0.$\!\!^{\circ}$13). 
While searching for gamma-ray emission from 3C 397, \citet{Bh17} and \citet{Er18} reported the detection of 3FHL J1907.0+0713/4FGL J1906.9+0712 within the energy range of 10 $-$ 300 GeV using more {\it Fermi}-LAT data. However, 3C 397 could not be detected in this study. 

The combined X-ray analysis of {\it ROSAT}, {\it ASCA} and {\it RXTE} data taken from 3C 397 showed that the overall spectrum is heavily absorbed and thermal in nature \citep{Sa00}. Due to the thermal nature of the X-ray emission that is located within its associated radio shell, 3C 397 is accepted as a mixed morphology (MM) SNR and its age was found to be approximately 5300 yr \citep{Sa05}. 

Using {\it Suzaku} data collected from 3C 397, \citet{Ya15} detected high abundances of stable Fe-peak elements (i.e. Ni and Mn), which indicate that this remnant is formed by the explosion of a white dwarf close to Chandrasekhar mass. Most studies based on the analysis of Fe K-shell emission suggested a Type Ia origin for 3C 397 \citep{Ch99, Ya13, Ya14}. \citet{Ya15} used {\it Spitzer} infrared data to calculate the ambient density for this remnant which was found to be relatively low confirming the Type Ia origin of 3C 397. \citet{MR20} showed that 3C 397 is likely the result of an energetic Type Ia explosion in a high-density ambient medium and that the progenitor was a near Chandrasekhar mass white dwarf.
 
 \citet{Ji10} reported that 3C 397 is confined within a pocket of molecular gas in the 27 $-$ 35 km s$^{-1}$ velocity interval except the south-east region of the SNR. The broadened $^{12}$CO($J$ = 1$-$0) line profile at about 32 km s$^{-1}$ is a strong kinematic evidence for the SNR shock-MC interaction \citep{Ji10, Kil16}. This MC is also resolved in $^{13}$CO($J$ = 1$-$0) and together with 3C 397 the 32 km s$^{-1}$ velocity component is found to be at the location of about 10.3 kpc \citep{Ca75,Ji10}. 
There is another CO component peaking at roughly 38 km s$^{-1}$, which in the 35 $-$ 42 km s$^{-1}$ velocity interval appears to form a crescent strip that is partially surrounding the western and southern borders of the SNR. This MC was suggested to be in connection to 3C 397 \citep{Sa05}, but it is likely to be a foreground MC toward the SNR at 2.1 kpc \citep{Ji10}, which probably causes the variations in the X-ray absorption from west to east \citep{Sa05} of the SNR.  

There is an H\,{\sc ii} region\citep[G41.1-0.2/G41.09-0.18;][]{Av02} located about 5 $-$ 7 arc-min to the west of the SNR, which was first separated from 3C 397 by \citet{Ca75}. Early studies located this region at the foreground of the SNR with a distance estimation of 3.6 and 9.3 kpc \citep{Ce90,Sa05}. \citet{Le16} reported the far-side distance of the H\,{\sc ii} region to be 10.5 $\pm$ 0.3 kpc for a constant rotational velocity and 9.8 $\pm$ 0.3 kpc for a linear rotational velocity at 58 km s$^{-1}$. So, although the H\,{\sc ii} region seems to be close to 3C 397 there is no clear evidence linking G41.1-0.2 to this SNR. However, G41.1-0.2 is related to the H\,{\sc ii} region C41.1-0.21 \citep{An09} at a local standard of rest velocity found to be about 59.9 km s$^{-1}$, as well as to G041.126-00.232 from the WISE catalog of H\,{\sc ii} regions \citep{An14} with a corresponding molecular ($^{13}$CO) local standard of rest velocity found to be about 60.2 km s$^{-1}$ \citep{Ur08}. 

Another H\,{\sc ii} region close to 3C 397 and 3FHL J1907.0+0713/4FGL J1906.9+0712 is G41.23-0.19 \citep{Av02}, which is an intermediate-mass SFR \citep{Lu14} also known as IRAS19049+0712 that is classified as 'Blob/Shell'-type by \citet{Lu14} showing an extended emission at 12 and 22 $\mu$m, including isolated blobs, shells, and enhancements at the edges of pillars or bright-rimmed clouds. The corresponding $^{13}$CO($J$ = 1$-$0) local standard of rest velocity was measured to be 59.2 km s$^{-1}$ with a near and far distance of 3.7 and 8.9 kpc, respectively \citep{Lu15}.  

At TeV energies, H.E.S.S. Galactic Plane Survey (HGPS) significance map showed no detection (roughly 3$\sigma$) at the locations of 3C 397 and 3FHL J1907.0+0713/4FGL J1906.9+0712. The upper limit at a predefined confidence level of 95 per cent was given as 2.12 $\times$ 10$^{-13}$ cm$^{-2}$ s$^{-1}$ \citep{Ab18}. 

In this paper, we analysed about 10 years of $\it Fermi$-LAT data to understand the origin of the gamma-ray emission arising from 3FHL J1907.0+0713/4FGL J1906.9+0712, which is positioned between PSR J1906+0722/4FGL J1906.4+0723 and 3C 397. Following the $\it Fermi$-LAT data reduction in Section \ref{Section2}, we applied the gamma-ray background model in Section \ref{Section3.1} and pulsar gating to the gamma-ray data using the pulsar PSR J1906+0722 ephemeris in Section \ref{Section3.2}. In Section \ref{Section3.3}, we further investigated the off-pulse gamma-ray emission and analysed the MC data taken by NANTEN2 in Section \ref{Section4}. In Section \ref{Section5}, we modelled the gamma-ray energy spectrum of 3FHL J1907.0+0713/4FGL J1906.9+0712 and showed the results of the fits. The results including the spatial correlation between the gamma-ray TS maps and CO intensity maps, as well as the dominating gamma-ray emission mechanisms are discussed in Section \ref{Section6}. Conclusions are presented in Section \ref{Section7}.

\section{Observations and Data Reduction}
\label{Section2}
The gamma-ray observations were taken from 2008-09-01 to 2019-02-04. In this analysis, we made use of the analysis packages \texttt{fermitools}\footnote{\url{http://fermi.gsfc.nasa.gov/ssc/data/analysis/software}} version \texttt{1.0.1} and  \texttt{fermipy}\footnote{\url{http://fermipy.readthedocs.io/en/latest/index.html}} version \texttt{0.17.4}. Using \texttt{gtselect} of \texttt{fermitools} we selected {\it Fermi}-LAT Pass 8 `source' class and `front$+$back' type events coming from zenith angles smaller than 90$^{\circ}$ and from a circular region of interest (ROI) with a radius of 20$^{\circ}$ centred at the location of R.A.(J2000) = 286$^{\circ}\!\!$.67 and decl.(J2000) = 7$^{\circ}\!\!$.33\footnote {This is the location of 3FGL J1906.6$+$0720 from the 3rd {\it Fermi}-LAT point-source catalog corresponding to PSR J1906$+$0722/4FGL J1906.4$+$0723}. The {\it Fermi}-LAT instrument response function version {\it P8R3$_{-}$SOURCE$_{-}\!\!$V2} was used. For mapping and morphological studies within the analysis region, events having energies in the range of 1 $-$ 300 GeV were selected. To deduce the spectral parameters of 3FHL J1907.0+0713/4FGL J1906.9+0712, events with energies between 200 MeV and 300 GeV were chosen.

\section{Analysis and Results}
\label{Section3}

\subsection{The background model}
\label{Section3.1}
In the background model, all the sources from 4FGL catalog within the ROI, as well as the galactic ({\it gll$_{-}$iem$_{-}$v07.fits}) and the isotropic ({\it iso$_{-}$P8R3$_{-}$SOURCE$_{-}\!\!$V2$_{-}\!\!$v1.txt}) diffusion components were included. With \texttt{gtlike}, we performed the maximum likelihood \citep{Ma96} fitting on data that was binned within the selected spatial and energy ranges. During the fit, the normalisation parameters of all the sources within 3$^{\circ}$ ROI, as well as the diffuse emission components were left free. Specifically, we freed the normalisation parameter of all sources with significance\footnote{The detection significance value is approximately equal to the square root of the test statistics (TS) value. Larger TS values indicate that the null hypothesis (maximum likelihood value for a model without an additional source) is incorrect.} $>$ 20 and we fixed all parameters for sources with significance $<$ 20. The parameters of all the other sources were fixed to values given in the 4FGL catalog. 

\begin{table}
 \caption{Parameters of PSR J1906+0722 as taken from \citet{Cl15}.}
 \begin{minipage}{100mm}
 \renewcommand{\arraystretch}{1.5}
    \begin{tabular}{@{}p{4.4cm}p{4.3cm}@{}}
		\hline \hline
		{\bf Parameter} & {\bf Value} \\ 
		\hline\hline
        Range of Photon Data (MJD)&  54710 $-$ 57902\\
	    
	    Reference epoch (MJD) & 55555 \\
     	
     	R.A.(J2000)  & 19$^h$ 06$^m$ 31$^s\!$.18 \\  
    	
    	Decl.(J2000) & $-$07$^{\circ}$ 22$'$ 55$''\!$.97 \\
        
        Frequency $f$ (Hz) &  8.9667089378363282748 \\
        
        1st freq. derivative $\dot{f}$ (Hz s$^{-1}$) &-2.8866285062981201978$\times\rm{10^{-12}}$ \\
        
        Glitch epoch (MJD) & 55066.2269630517\\
        
        Per. $\!f$ glitch increment $\!\triangle f$ (Hz) & 4.0329282031051799999$\times\rm{10^{-5}}$\\
        
        Per. $\!\dot{f}$ glitch increment $\!\!\triangle \dot{f}$ (Hz s$^{-1}$) & -2.51228058941666$\times\rm{10^{-14}}$\\
        
        Dec. $\!f$ glitch increment $\triangle f_{d}$ (Hz) & 3.57557668242071$\times\rm{10^{-7}}$ \\
        
        Glitch decay time const. $\tau_{d}$ (days) & 221.628841809732\\
        \hline \hline
	\end{tabular}
    \label{table_1}
\end{minipage}
\end{table} 

\begin{figure*}
     \includegraphics[width=1.0\textwidth]{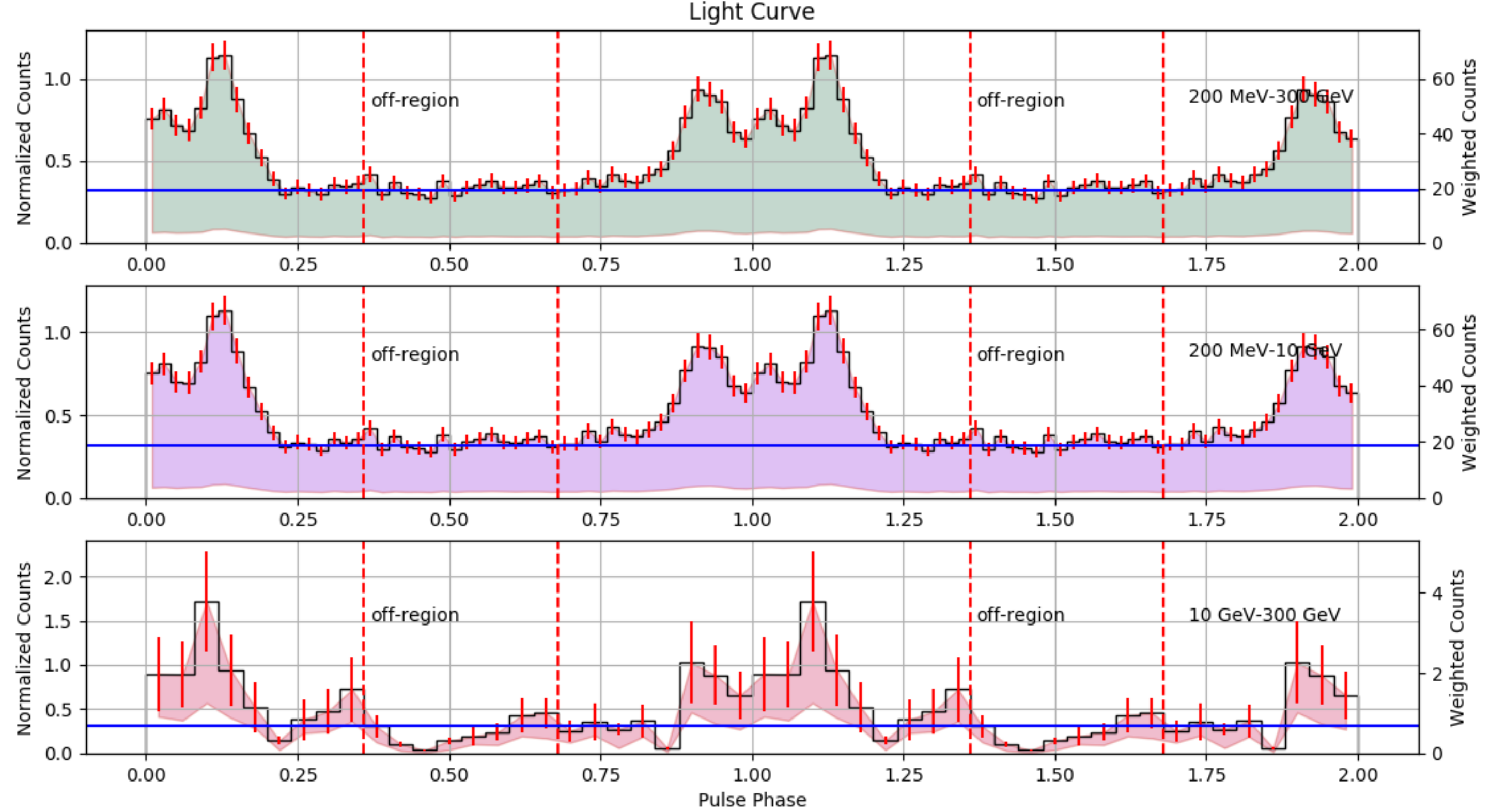}
    \caption{Weighted light curve of PSR J1906+0722 for three energy ranges (upper panel: 0.2 $-$ 300 GeV, middle panel: 0.2 $-$ 10 GeV, lower panel: 10 $-$ 300 GeV). The blue line in each panel denotes the background level and red error bar defines the statistical errors in the light curve in each phase bin. }
    \label{figure_1}
    \vspace{-0.5cm}
\end{figure*}

\subsection{Pulsar gating and pulsar light curve}
\label{Section3.2}
We applied the pulsar gating technique (PGT)\footnote{\url{https://fermi.gsfc.nasa.gov/ssc/data/analysis/scitools/pulsar_gating_tutorial.html}} to remove the pulsar PSR J1906+0722 from our analysis region so that we can analyse the rest of the gamma-ray emission. We used an updated ephemeris for this pulsar \citep{Cl15} shown in Table \ref{table_1}, to assign pulse phases with {\it Fermi}-LAT plug-ins \citep{Ra11} of \texttt{TEMPO2} \citep{Ho06}. 

In order to define the on- and off-pulse intervals in phase space, we produced a pulsar light curve. The light curve is produced using the {\it Fermi}-LAT Aperture Photometry\footnote{\url{https://fermi.gsfc.nasa.gov/ssc/data/analysis/scitools/aperture_photometry.html}}. For the event weighting with \texttt{gtsrcprob}, we calculated the photon weights on all the photons within 2$^{\circ}$ ROI around the position of PSR J1906$+$0722/4FGL J1906.4$+$0723.

Fig.  \ref{figure_1} shows the phase folded light curve for two rotation periods originated from the weighted photons for three energy ranges. The upper panel shows the light curve for full energy range (200 MeV to 300 GeV) and the other two light curves are generated for lower and higher energies. The phase folded light curves show the same pulsed nature as we obtained from \citet{Cl15} and for making our analysis conservative, we have selected the same off-pulse phase interval as obtained by \citet{Cl15}. In Fig. \ref{figure_1}, for one single rotation, we have denoted phase 0.36 to 0.68 as off-pulse phase and the rest phase as on-pulse phase interval. 

\begin{figure*}
\includegraphics[width=1.0\textwidth]{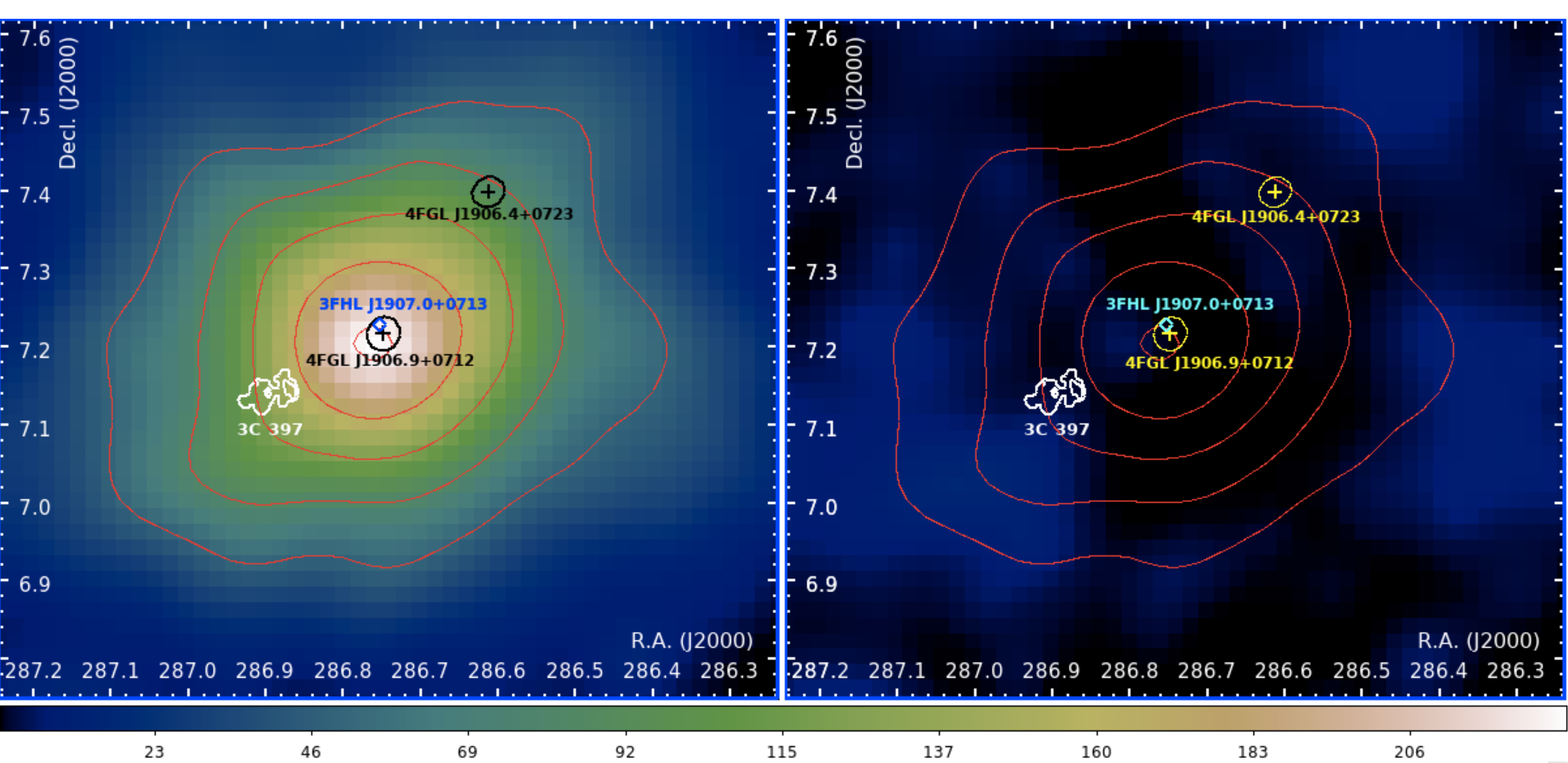}
\caption{TS map of off-pulse gamma-ray emission of the 3FHL J1907.0+0713/4FGL J1906.9+0712 analysis region in the energy range of 1 $-$ 300 GeV. On both left and right panels, 3C 397 {\it ROSAT} X-ray contours are indicated (white contours). In both panels, red contours are the TS levels for gamma rays at 49, 81, 121, 169, 225. The positions of the {\it Fermi}-LAT point sources are shown with plus markers and their positional errors are shown with ellipses. The source previously detected by the 3FHL source catalog is shown with a diamond marker. Left Panel: The background gamma-ray emission model does not include 3FHL J1907.0+0713/4FGL J1906.9+0712. Right Panel: The background gamma-ray emission model includes 3FHL J1907.0+0713/4FGL J1906.9+0712.}
\label{figure_2}
\end{figure*}

\subsection{Off-pulse gamma rays}
\label{Section3.3}
To obtain the off-pulse gamma-ray distribution, any possible contamination due to the strong emissions from the pulsar itself has to be removed. So, after assigning the phase periods, we used the \texttt{ftselect} of the HEAsoft analysis package\footnote{\url{https://heasarc.gsfc.nasa.gov/docs/software/heasoft/}} to separate the on- and off-pulse data sets from each other. Then, we continued with the \texttt{fermitools} binned likelihood analysis\footnote{\url{https://fermi.gsfc.nasa.gov/ssc/data/analysis/scitools/binned_likelihood_tutorial.html}} on the extracted off-pulse data. 

\vspace{-0.5cm}
\subsubsection{Source morphology}
\label{Section3.3.1}
The left panel of Fig. \ref{figure_2} shows the initial TS map produced in the energy range of 1 $-$ 300 GeV for the 10$^{\circ}$ $\times$ 10$^{\circ}$ analysis region. In this map, all point-sources from the 4FGL catalog, except 4FGL J1906.9+0712, are included in the gamma-ray background model. In this case, the gamma-ray emission peaks at the position of 4FGL J1906.9+0712. 

We computed the best-fitting position of the gamma-ray excess to be (RA, decl.) = (286$^{\circ}\!\!$.77 $\pm$ 0$^{\circ}\!\!$.02, 7$^{\circ}\!\!$.19 $\pm$ 0$^{\circ}\!\!$.01) by utilising the \texttt{localize()} algorithm of \texttt{fermipy}. The best-fitting position overlaps with the previously reported locations of 3FHL J1907.0+0713\footnote{(RA, decl.) = (286$^{\circ}\!\!$.7524 $\pm$ 0$^{\circ}\!\!$.0463, 7$^{\circ}\!\!$.2287 $\pm$ 0$^{\circ}\!\!$.0463)} and 4FGL J1906.9+0712\footnote{(RA, decl.) = (286$^{\circ}\!\!$.6115 $\pm$ 0$^{\circ}\!\!$.0186, 7$^{\circ}\!\!$.3982 $\pm$ 0$^{\circ}\!\!$.0214)} at a 95 per cent confidence level. Fig. \ref{figure_2} right panel shows the TS map for which the background gamma-ray emission model includes 3FHL J1907.0+0713/4FGL J1906.9+0712 as a point-like source. 

The \texttt{extension} method of \texttt{fermipy} calculates the likelihood ratio between the point-source hypothesis and a best fit extension model. The best-fitting extension is found by performing a likelihood profile scan over the source width. The extension of 3FHL J1907.0+0713/4FGL J1906.9+0712 was tested with the {\it Radial Gaussian} extension model. This measurement resulted in this source being a point-like source with TS$_{\rm ext}$\footnote{The TS of the extension (TS$_{\rm ext}$) parameter is the likelihood ratio that compares the likelihood for being a point-like source ($L_{\rm pt}$) to a likelihood for an existing extension ($L_{\rm ext}$), TS$_{\rm ext}$ = -2log($L_{\rm ext}$/$L_{\rm pt}$). A gamma-ray source's extension is detected for TS$_{\rm ext}~\geqslant$25.} being less than 25.

Since multiple sources, 3FHL J1907.0+0713/4FGL J1906.9+0712, PSR J1906+0722/4FGL J1906.4+07234 and 3C 397, are located very close to each other, it is important to understand how these sources are contributing to the gamma-ray emission that we observe within the analysis region. So, to find out the source combination that best describes the gamma-ray emission, the following multiple source models were defined in the gamma-ray background model and fit to the data, where the prefix {\it PS} stands for {\it Point Source} : 
\begin{itemize}
\item {\it 3PS Model} includes 3FHL J1907.0+0713/4FGL J1906.9+0712, 3C 397, and PSR J1906+0722/4FGL J1906.4+07234 assuming all having power-law (PL) type spectrum,
\item {\it 2PS Model-1} includes only 3FHL J1907.0+0713/4FGL J1906.9+0712 and 3C 397 assuming both having PL type spectrum,
\item {\it 2PS Model-2} includes 3FHL J1907.0+0713/4FGL J1906.9+0712 and PSR J1906+0722/4FGL J1906.4+0723 assuming both having PL type spectrum,
\end{itemize}

We implemented the Akaike Information Criterion (AIC) \citep{Ak98,La12} for each model to eventually select the best-fitting multi-source model. AIC is given by the following equation:
\begin{equation}
 \\
 AIC=2k-2ln(L), 
\end{equation}
where {\it k} is the number of estimated parameters in the model and $L$ is the maximum value of the likelihood function for the model. The best source model is considered to be the one that minimises the AIC value. So, $\Delta$AIC = (AIC)$_{1}$ $-$ (AIC)$_{m}$ is used to compare the models, which are the {\it 3PS Model} with {\it k} = 6 (given by the index value 1), with other two {\it 2PS Models} with index {\it m} = 1, 2 (both having {\it k} = 4) tested in this analysis. 

\begin{table}
    \caption{Multi-source spatial fit results found for the 1 $-$ 300 GeV energy range. Column (1) shows the spatial model name. (2)nd and (3)rd columns give the degrees of freedom (d.o.f.) of each model and the corresponding $\Delta$AIC value, respectively. }
    \begin{minipage}{0.45\textwidth}
    	\centering
	\begin{tabular}{lccc} 
        \hline\hline
        Spatial Model                                             &d.o.f.      &$\Delta$AIC   \\
        $~~~~~~$(1)                                              &(2)          &(3)   \\ 
 		\hline
           \\
         \multirow{1}{*}{\it{2PS Model-2}}               &4            &0   \\
         \\
         \multirow{1}{*}{\it{2PS Model-1}}               &4            &0.2 \\
         \\                 
         \multirow{1}{*}{\it{3PS Model}}                 &6            &2   \\
        \hline 
        \label{table_2}
        \vspace{-0.3cm}
	\end{tabular}
    \end{minipage}
\end{table}

Table \ref{table_2} shows the fit results, where the best fitting multiple source model is {\it 2PS Model-2}, in which case we added 3FHL J1907.0+0713/4FGL J1906.9+0712 and PSR J1906+0722/4FGL J1906.4+0723 into the gamma-ray background model assuming both sources as point-like in nature with PL-type spectra. It should also be noted that in none of these spatial model fits (i.e. 2PS Model-2, 2PS Model-1 and 3PS Model), neither 3C 397 nor 4FGL J1906.4+0723 had a significance above 4$\sigma$. 

\vspace{-0.5cm}
\subsubsection{Testing variability}
\label{Section3.3.2}
\begin{figure}
\includegraphics[width=0.5\textwidth]{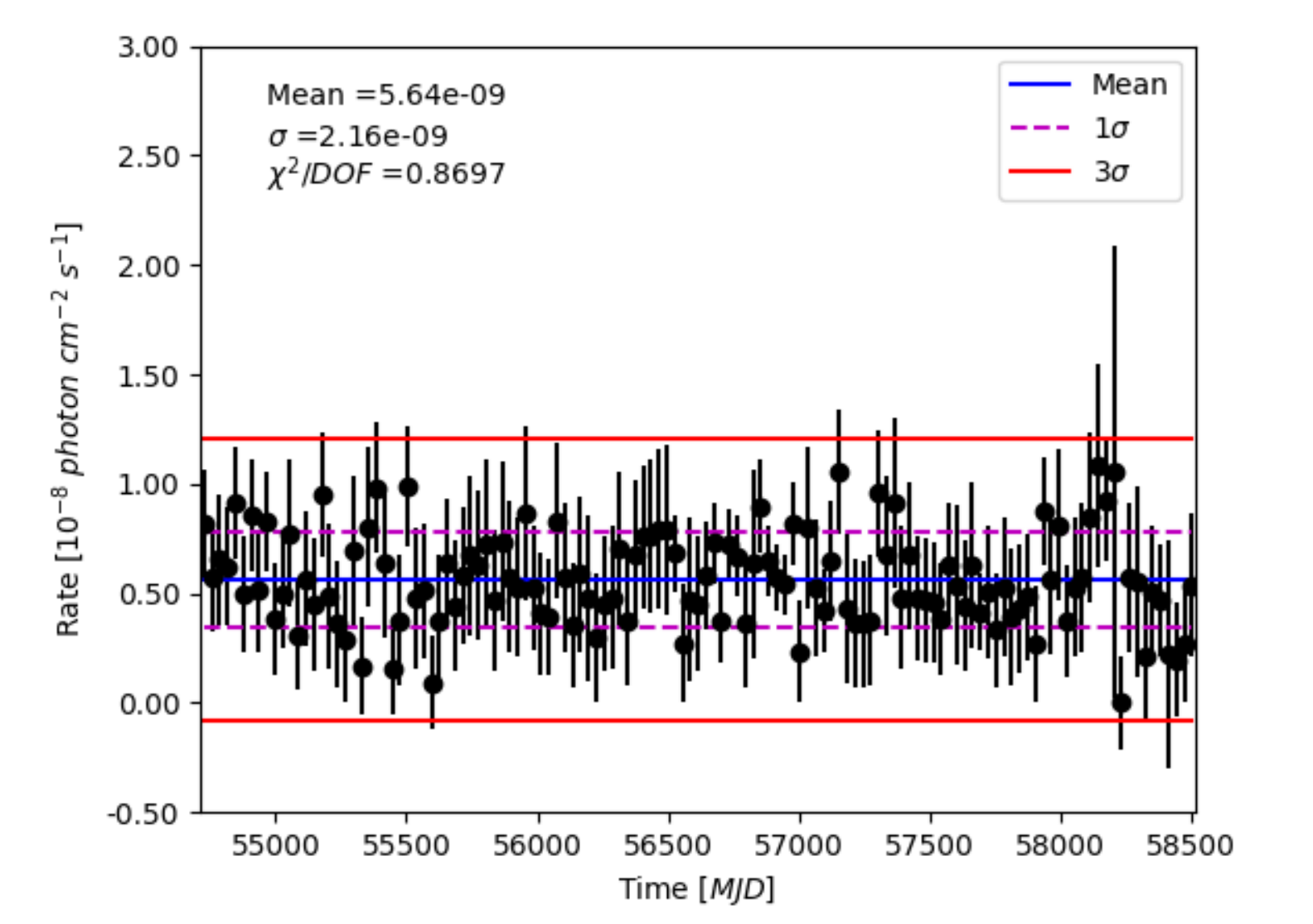}
\caption{1-month-binned gamma-ray light curve extracted at the location of 3FHL J1907.0+0713/4FGL J1906.9+0712 using the off-pulse gamma-ray data in the energy range of 200 MeV $-$ 300 GeV.} 
\label{figure_3}
\end{figure}

We first looked for long term variability in the light curve of 3FHL J1907.0+0713/4FGL J1906.9+0712 by taking data from the circular region of 1$^{\circ}$ around the best-fitting position.
Fig.  \ref{figure_3} shows the 1-month binned light curve obtained after applying {\it Fermi}-LAT {\it aperture photometry}, where we checked for possible variations in the flux levels. If any or some of the flux data points are above 3$\sigma$, this would be an indication of a significant variability (e.g. flare) in the circular region of interest. In Fig. \ref{figure_3}, all of the flux data points remain within the 1$\sigma$ and 3$\sigma$ bands. Thus, we conclude that there is no long term variability observed in the close neighbourhood.

\begin{table*}
    \caption{$\Delta$AIC results for different spectral models combinations used for the pulsar and 3FHL J1907.0+0713/4FGL J1906.9+0712 produced using the off-pulse data in the energy range of 0.2 $-$ 300 GeV. Column (1) shows the spectral model name. (2)nd and (3)rd columns give the d.o.f. of each model and the corresponding $\Delta$AIC value, respectively. }
    \centering
    \begin{minipage}{1.0\textwidth}
	\begin{tabular}{lccccc} 
        \hline\hline
        Spectral Model Name       & 4FGL J1906.4+0723/PSR J1906+0722  & 3FHL J1907.0+0713/4FGL J1906.9+0712  & Total d.o.f. &$\Delta$AIC   \\
        $~~~~~~$(1)                     & (2)                & (3)               & (4)         & (5)           \\ 
 		\hline
         \\
         \multirow{1}{*}{Model-3}   & LP                 & PL                 &5      &0   \\
         \\
         \multirow{1}{*}{Model-4}   & LP                 & LP                 &6      &53.14    \\
         \\                 
         \multirow{1}{*}{Model-1}   & PL                 & PL                 &4      &67.16 \\
         \\             
         \multirow{1}{*}{Model-2}   & PL                 & LP                 &5      &132.74    \\
         \\                 
         \multirow{1}{*}{Model-6}   & PLSEC2             & LP                 &6      &146.50    \\
         \\  
          \multirow{1}{*}{Model-9}  & LP                 & PLEC               &6      &206.32    \\ 
         \\
         \multirow{1}{*}{Model-5}   & PLSEC2             & PL                 &5      &331.64    \\
         \\                 
          \multirow{1}{*}{Model-8}  & PL                 & PLEC               &6      &474.09    \\ 
         \\
          \multirow{1}{*}{Model-7}  & PLSEC2             & PLEC               &7      &1305.04    \\ 
        
        \hline 
        \label{table_3}
        \vspace{-0.3cm}
	\end{tabular}
    \end{minipage}
\end{table*}

\subsubsection{Spectral energy distribution}
\label{Section3.3.3}
The spectral model for 4FGL J1906.4+0723/PSR J1906+0722 in the 4FGL source catalog is the power-law super exponential cutoff (PLSuperExpCutoff2; PLSEC2) model as given below: 
\[
~~~~~~~~~~~~~~~~~~~\mbox{dN/dE} = \mbox{N}_0~\left( \mbox{E}/\mbox{E}_0 \right)^{\Gamma_1} \mbox{exp}(-\mbox{p}_{1}\mbox{E}^{\Gamma_2})
\]
where ${\rm N_0}$ is the prefactor, ${\rm \Gamma_1}$ and ${\rm \Gamma_2}$ are spectral indices, ${\rm E_0}$ is the scale and p$_1$ is the exponential factor. We also tested the following spectral models\footnote{\url{https://fermi.gsfc.nasa.gov/ssc/data/analysis/scitools/source_models.html}} for 3FHL J1907.0+0713/4FGL J1906.9+0712:
 \[
\mbox{PowerLaw (PL):}~~~~\mbox{dN/dE} = \mbox{N}_0~(\mbox{E}/\mbox{E}_0)^{{\rm \Gamma}}
  \]
\vspace{-0.5cm}
 \[
\mbox{LogParabola (LP):}~\mbox{dN/dE} = \mbox{N}_0 ~(\mbox{E}/\mbox{E}_b)^{{\rm-(\alpha + \beta \log(E/E_{b}))}}
 \]
 \vspace{-0.5cm}
 \[
\mbox{ExpCutoff (PLEC):}\mbox{dN/dE} = \mbox{N}_0(\mbox{E}/\mbox{E}_0)^{{\rm \Gamma}}\exp{(-(\mbox{E}-\mbox{E}_{b})/{\rm p_1})} 
 \]
where ${\rm E_0}$ is the scale, E$_b$ is the break parameter, and ${\rm p_1}$ is the exponential factor. $\Gamma$ and (${\rm \beta}$, ${\rm \alpha}$) are spectral indices of the PL/PLEC and LP spectral model, respectively. N$_0$ is the prefactor.

We made use of the AIC method \citep{Ak98,La12} again, this time for finding out the best fitting combination of spectral models for the case of {\it 2PS Model-2} spatial model. We used the above-mentioned three spectral models (i.e. PLSEC2, PL and LP) for 4FGL J1906.4+0723/PSR J1906+0722 and three spectral models (i.e. PL, LP and PLEC) for 3FHL J1907.0+0713/4FGL J1906.9+0712. For each spectral model the {\it k} parameter is taken as 2 for the PL model, 3 for the LP and PLSEC2 models, and 4 for the PLEC model.

The $\Delta$AIC values are tabulated in Table \ref{table_3} for different models (i.e. Model-1 to Model-9). The lowest $\Delta$AIC points the best-fitting spectral model combination, which is Model-3. However, the second spectral parameter ($\beta$ parameter) for the LP spectral model for PSR J1906+0722/4FGL J1906.4+0723 was not fit properly, where it was found to be approximately 10$^{-5}$. This means that the fit was not successfully completed, where the first and second sources (3FHL J1907.0+0713/4FGL J1906.9+0712 and PSR J1906+0722/4FGL J1906.4+0723) seem to be correlated with each other in the fit. So, we look at Model-4, which has the second lowest  $\Delta$AIC values. This model did not fit correctly, too, where the TS value of 3FHL J1907.0+0713/4FGL J1906.9+0712 was found below 25, which contradicts with the best-fitting spatial model, {\it 2PS Model-2}. Finally, we switched to the next best-fitting model with the lowest $\Delta$AIC value, Model-1. 

In the 200 MeV $-$ 300 GeV energy range, the results for the best-fitting spectral model combination (i.e. Model-1) are given below: 
\begin{itemize}
\item 3FHL J1907.0+0713/4FGL J1906.9+0712 was detected with a TS value of 178 (about 13$\sigma$). The PL-spectral index was found to be $\Gamma$=2.26 $\pm$ 0.05. The total flux and energy flux values were found to be (1.03 $\pm$ 0.11)$\times$10$^{-8}$ cm$^{-2}$ s$^{-1}$ and (8.51 $\pm$ 0.78)$\times$10$^{-6}$ MeV cm$^{-2}$ s$^{-1}$, respectively.
\vspace{0.3cm}
\item PSR J1906+0722/4FGL J1906.4+0723 was detected with a TS value of 204 (about 14$\sigma$). The PL-type spectral fit resulted in ${\Gamma}$ = 2.65 $\pm$ 0.06 for the spectral index. The photon and energy fluxes were found to be (1.74 $\pm$ 0.14)$\times$10$^{-8}$ cm$^{-2}$ s$^{-1}$ and (8.71 $\pm$ 0.66)$\times$10$^{-6}$ MeV cm$^{-2}$ s$^{-1}$, respectively. 
\end{itemize}

The systematic uncertainties of the spectral data points are evaluated by altering the normalization of the Galactic diffuse emission by $\pm$6\%, running \texttt{gtlike} for each energy bin, and evaluating the change in flux \citep{Ab09}, which are shown on the plotted energy spectra in Figure \ref{figure_7}.
 
\section{Molecular Clouds}
\label{Section4}
\subsection{Observations and data reduction}
\label{Section4.1}
Observations of $^{12}$CO($J$ = 1$-$0) emission line at 2.3 mm wavelength were conducted in October and December 2012 using the NANTEN2 4 m mm/sub-mm radio telescope of Nagoya University located at Pampa La Bola in northern Chile (approximately 4865 m above sea level). We observed an area of 2$^{\circ}\times\!\!$ 2$^{\circ}$ around the SNR 3C 397 by using the on-the-fly mapping mode with a Nyquist sampling. A 4 K cooled Nb superconductor-insulator-superconductor (SIS) mixer receiver was used as the front-end. The back-end was a digital Fourier-transform spectrometer with 16,384 channels of 1 GHz bandwidth. The velocity coverage and resolution are about 2,600 km s$^{-1}$ and 0.16 km s$^{-1}$, respectively. Typical system temperature including the atmosphere is about 150 - 200 K in the double side band (DSB). The absolute intensity calibration was done by observing IRAS 16293-2433 [($\alpha_\mathrm{J2000}$, $\delta_\mathrm{J2000}$) = ($16^\mathrm{h}32^\mathrm{m}23\fs3$, $-24^{\circ} 28\farcm 39\farcs 2$)] \citep{Ri06}. The pointing accuracy was within 15 arc-seconds. After convolution with a two-dimensional Gaussian kernel of 90 arc-seconds (full-width half-maximum; FWHM), we obtained the final data with the beam size of approximately 180 arc-seconds (FWHM). The typical noise fluctuation is about 0.43 K at the velocity resolution of about 0.95 km s$^{-1}$.

\begin{figure*}
\centering
\includegraphics[width=0.9\textwidth]{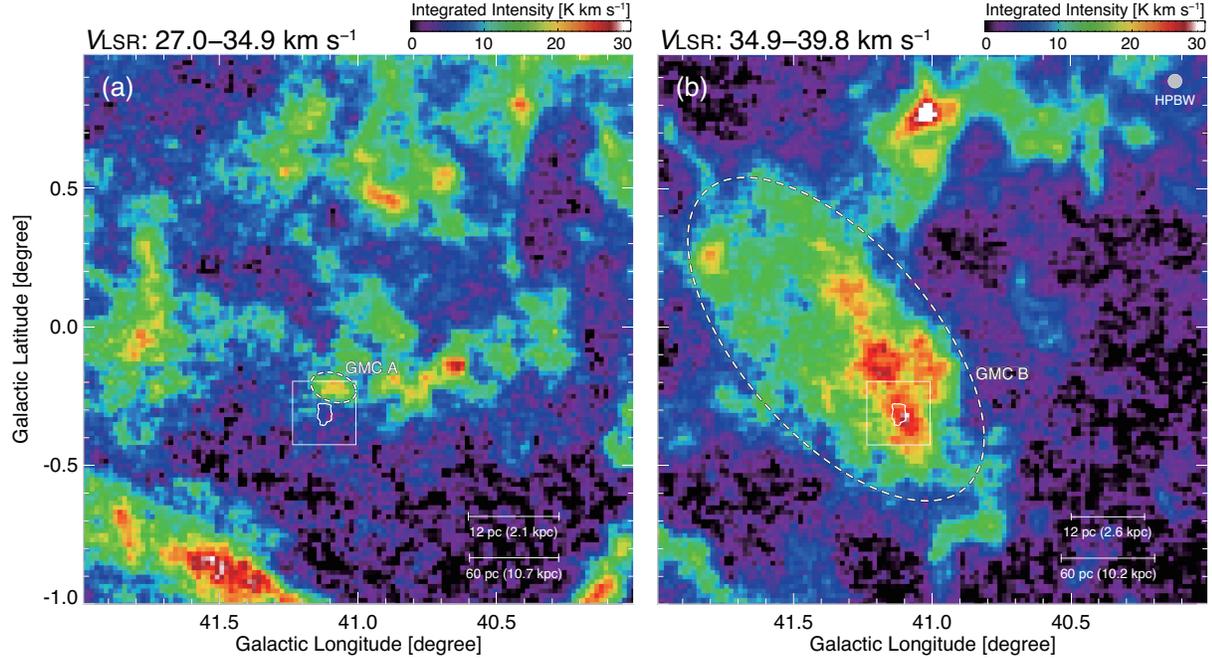}
\caption{The integrated $^{12}$CO($J$ = 1$-$0) intensity at 27.0 $-$ 34.9 km s$^{-1}$ (Left Panel) and 34.9 $-$ 39.8 km s$^{-1}$ (Right Panel) taken by NANTEN2, where the location of GMC A and GMC B is shown, respectively. The white frames laid on both of these maps are representing the regions shown in Fig. \ref{figure_5}. The solid contours indicate boundaries of Chandra X-ray shell, where the 0.3 - 2.1 keV image was taken from the Chandra Supernova Remnant Catalog (\url{https://hea-www.harvard.edu/ChandraSNR/G031.9+00.0/2786/work/acis_E300-2100_FLUXED_G2.fits.gz}) with the lowest contour level is at 0.9 $\times$ 10$^{-7}$ photons cm$^{-2}$ s$^{-1}$ pixel$^{-1}$.}
    \label{figure_4}
\end{figure*}

\begin{figure*}
\includegraphics[width=\textwidth]{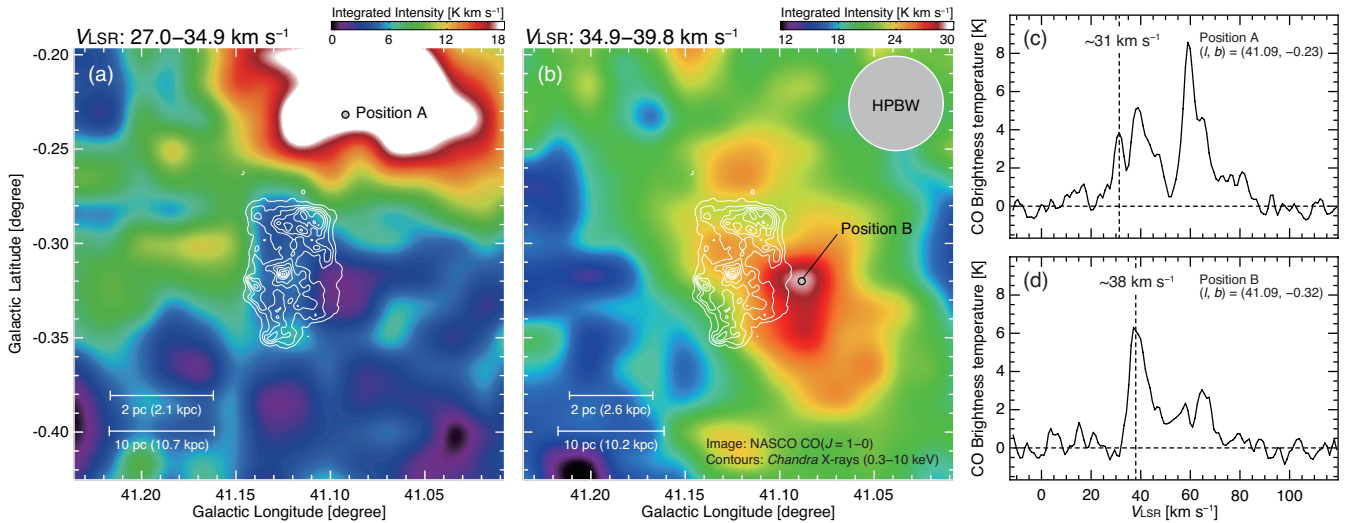}
\caption{The integrated intensity of NANTEN2 $^{12}$CO($J$ = 1$-$0) data collected from the neighbourhood of 3C 397. Panel (a) on the left and Panel (b) in the middle show the SNR in white contours ({\it Chandra} X-ray data). On panel (a) GMC A appears in the velocity range of 27.0 $-$ 34.9 km s$^{-1}$ with its position marked with \lq Position A\rq. On panel (b) GMC B appears in the velocity range of 34.9 $-$ 39.8 km s$^{-1}$ with its position marked with \lq Position B\rq. The half-power beam-width (HPBW) of the observation is shown by a grey-filled circle on panel (b). Panel (c) and panel (d) show the CO spectra toward the \lq Positions A and B\rq, where the peak velocity of GMC A is shown at about 31 km s$^{-1}$ and that of GMC B is shown at about 38 km s$^{-1}$.   }
    \label{figure_5}
\end{figure*}

\subsection{Analysis results}
\label{Section4.2}
Fig.  \ref{figure_4} shows the large-scale $^{12}$CO($J$ = 1$-$0) maps covering from galactic longitude, $l$ = 40$^{\circ}$ to 42$^{\circ}$ obtained with the NANTEN2 superposed on the {\it Chandra} X-ray boundary of 3C 397. For the velocity range of 27.0 $-$ 34.9 km s$^{-1}$ (Fig.  \ref{figure_4}a), a giant MC (GMC, hereafter refer to as GMC A) happens to be located on the northwest of the SNR which was previously mentioned by \citet{Ji10} and \citet{Kil16}. For the velocity range of 34.9 $-$ 39.8 km s$^{-1}$ (Fig.  \ref{figure_4}b), we find a large GMC (hereafter refer to as GMC B) which is elongated from southwest to northeast across 3C 397. We also note that GMC A has a clumpy distribution with a diameter of roughly 0.2$^{\circ}$, whereas GMC B shows a strongly elongated distribution of roughly 1.5$^{\circ}$, possibly indicating that GMCs A and B are located on the far-side and near-side of the galaxy, respectively.
\begin{table*}
	\centering
    \caption{Properties of GMC A and GMC B toward 3C 397 and 3FHL J1907.0+0713/4FGL J1906.9+0712.}
    \begin{minipage}{1.\textwidth}
	\begin{tabular}{l c c cccccccc} 
        \hline\hline
        Name &{\it l} &{\it b} &T$_{peak}$ &$V_\mathrm{peak}$ &$\Delta V$ &Size &Mass &Density &Distance &Comments \\
        
        $~$ &($^{\circ}$) &($^{\circ}$) &(K) &(km s$^{-1}$) &(km s$^{-1}$) &(pc) &(M$_{\odot}$) &(cm$^{-3}$) &(kpc) & $~$ \\
 (1) &(2) &(3) &(4) &(5) &(6) &(7) &(8) &(9) &(10) &(11) \\ [0.5ex]
 		\hline
        
        \multirow{2}{*}{GMC A} & \multirow{2}{*}{41.10} & \multirow{2}{*}{-0.22} & \multirow{2}{*}{4.2} & \multirow{2}{*}{31.1} & \multirow{ 2}{*}{4.5} & 5.8 & 1900 & 690 & 2.1 & near-side distance \\
         & & & & & & 29.8 & 49000 & 140 & 10.7 & far-side distance (adopted) \\
         
        \multirow{2}{*}{GMC B} & \multirow{2}{*}{41.10} & \multirow{2}{*}{-0.32} & \multirow{2}{*}{6.6} & \multirow{ 2}{*}{38.1} & \multirow{2}{*}{5.7} & 26.8 & 86000 & 330 & 2.6 & near-side distance (adopted)  \\
         & & & & & & 105.2 & 133000 & 80 & 10.2 & far-side distance \\
        \hline 
        \label{table_4}
         \vspace{-0.4cm}
	\end{tabular}
    \raggedright{{\bf Note:} Col. (1): GMC name. Cols. (2$-$7): Properties of $^{12}$CO($J$ = 1$-$0) emission line obtained by a single Gaussian fitting. Cols. (2)$-$(3): Position of peak intensity. Col. (4): Peak brightness temperature. Col. (5): Centre velocity. Col. (6): Line width (FWHM). Col. (7): GMC size defined as 2$\sqrt{(A/\pi)}$, where {\it A} is the total cloud surface area enclosed by the half intensity contours of the maximum integrated intensity. Col. (8): GMC mass derived by using a relation between the $^{12}$CO($J$ = 1$-$0) integrated intensity $W$(CO) and the molecular hydrogen column density $N$(H$_\mathrm{2}$) as $N$(H$_\mathrm{2}$) = 2.0 $\times$ 10$^{20}$[$W$(CO) (K km s$^{-1}$)] (cm$^{-2}$) \citep{Be93}. Col. (9): Number density of hydrogen in GMCs. Col. (10): Kinematic distance derived by a relation from \citet{Br93}.} \\
    \end{minipage}
\end{table*}

\begin{figure}
\includegraphics[width=0.45\textwidth]{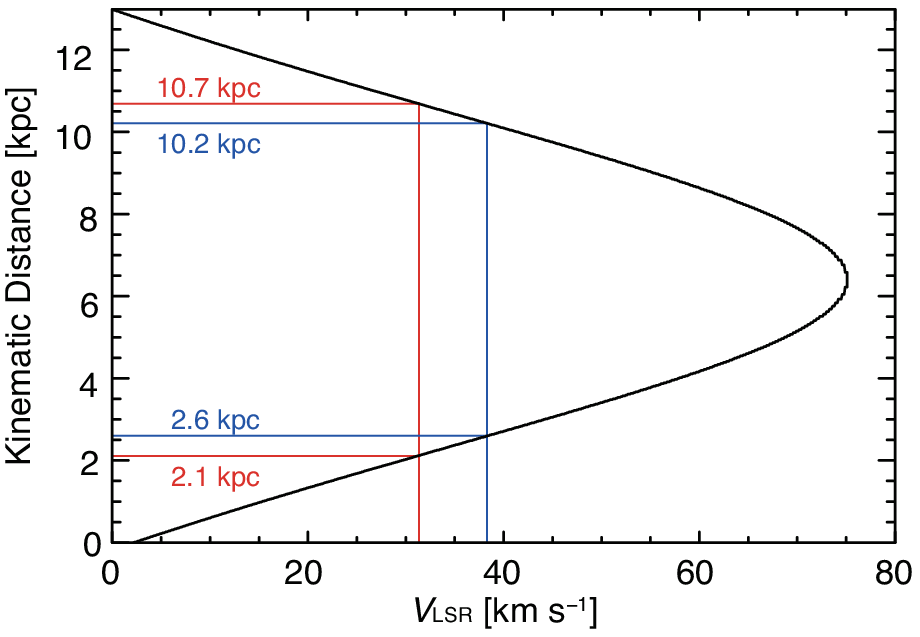}
\caption{The kinematic distance derived for GMC A is 2.1 kpc (near side) and 10.7 kpc (far side), and that of GMC B is 2.6 kpc (near side) and 10.2 kpc (far side). We accepted the physical properties of far-side distance (about 10.7 kpc) for GMC A and that of near-side distance (about 2.6 kpc) for GMC B.}
    \label{figure_6}
\end{figure}

Figs. \ref{figure_5}a and \ref{figure_5}b show the enlarged views of $^{12}$CO($J$ = 1$-$0) toward 3C 397 superposed on the X-ray contours of the SNR 3C 397. Diffuse MCs and GMC A are located nicely along with the SNR shell. On the other hand, we newly find that the bright $^{12}$CO($J$ = 1$-$0) peak of GMC B spatially coincides not only with the outer boundary of the SNR shell, but also with the GeV gamma-ray peak in the energy range from 200 MeV to 300 GeV. Figs. \ref{figure_5}c and \ref{figure_5}d show the $^{12}$CO($J$s = 1$-$0) spectra toward the positions A and B as shown in Figs. \ref{figure_5}a and \ref{figure_5}b. The peak velocity of GMC A is about 31 km s$^{-1}$ and that of GMC B is about 38 km s$^{-1}$. According to \citet{Br93}, a relation between the kinematic distance and radial velocity toward the positions A and B is shown as Fig. \ref{figure_6}. We therefore derive the kinematic distance of GMC A is approximately 2.1 kpc (near side) and 10.7 kpc (far side), and that of GMC B is approximately 2.6 kpc (near side) and 10.2 kpc (far side). Definitions and fundamental physical properties of each GMC and distance are listed in Table \ref{table_4}. In the present paper, we use the physical properties of far-side distance (about 10.7 kpc) for GMC A and that of near-side distance (about 2.6 kpc) for GMC B.

As indicated in Table \ref{table_4}, the number density of hydrogen in GMC A and GMC B were found to be 140 cm$^{-3}$ and 330 cm$^{-3}$ for the adopted distance of about 10.7 kpc and 2.6 kpc, respectively, which are consistent with the mean volume densities of molecular clouds in the Milky Way \citep{HeDa15} and they are much larger than the typical density of the interstellar medium (ISM) around the Sun (roughly 0.1 cm$^{-3}$) \citep{CoRe87}. 

\begin{figure*}
\begin{tabular}{cc}
	\includegraphics[width=0.49\textwidth]{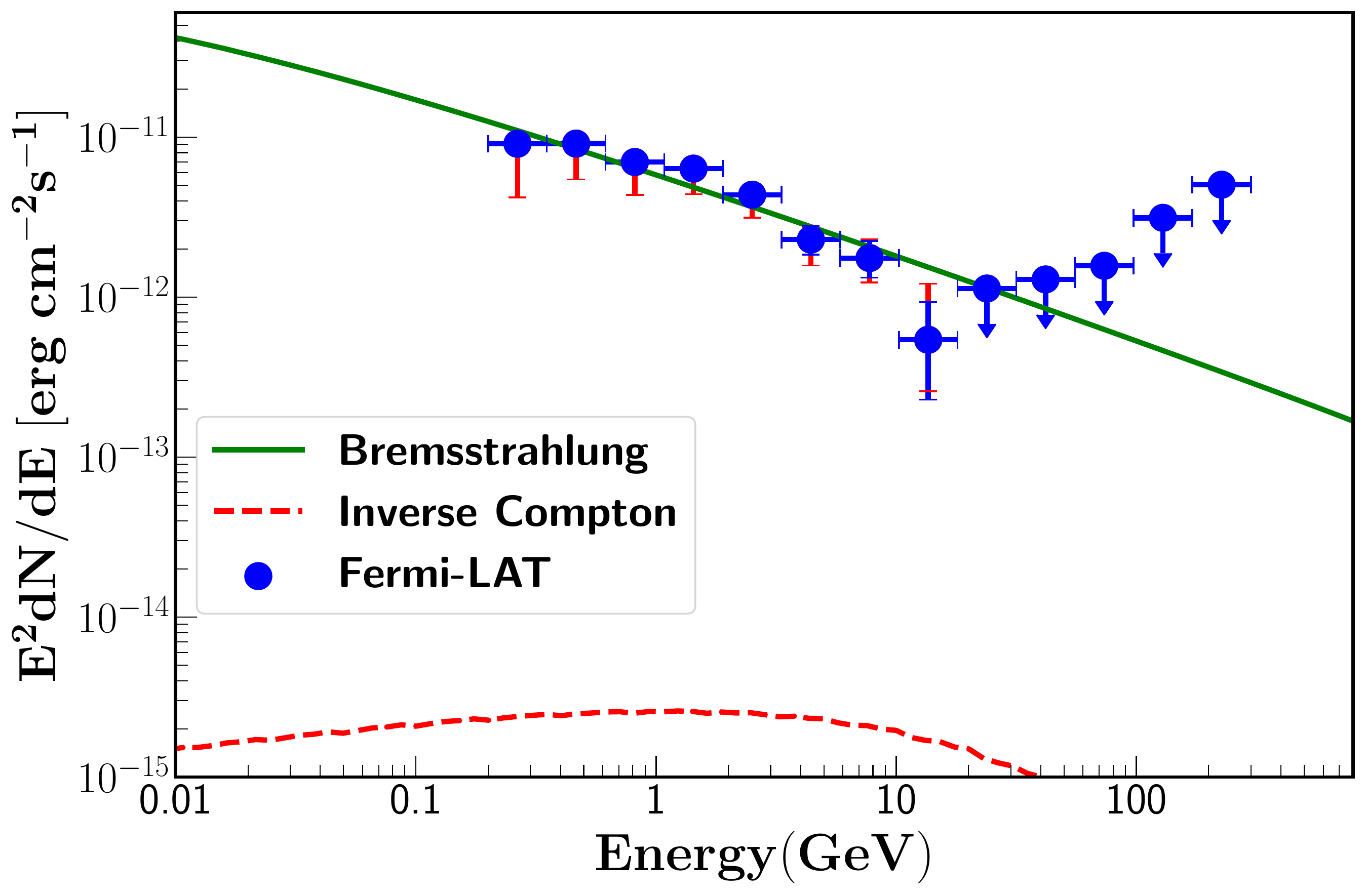}
        \includegraphics[width=0.49\textwidth]{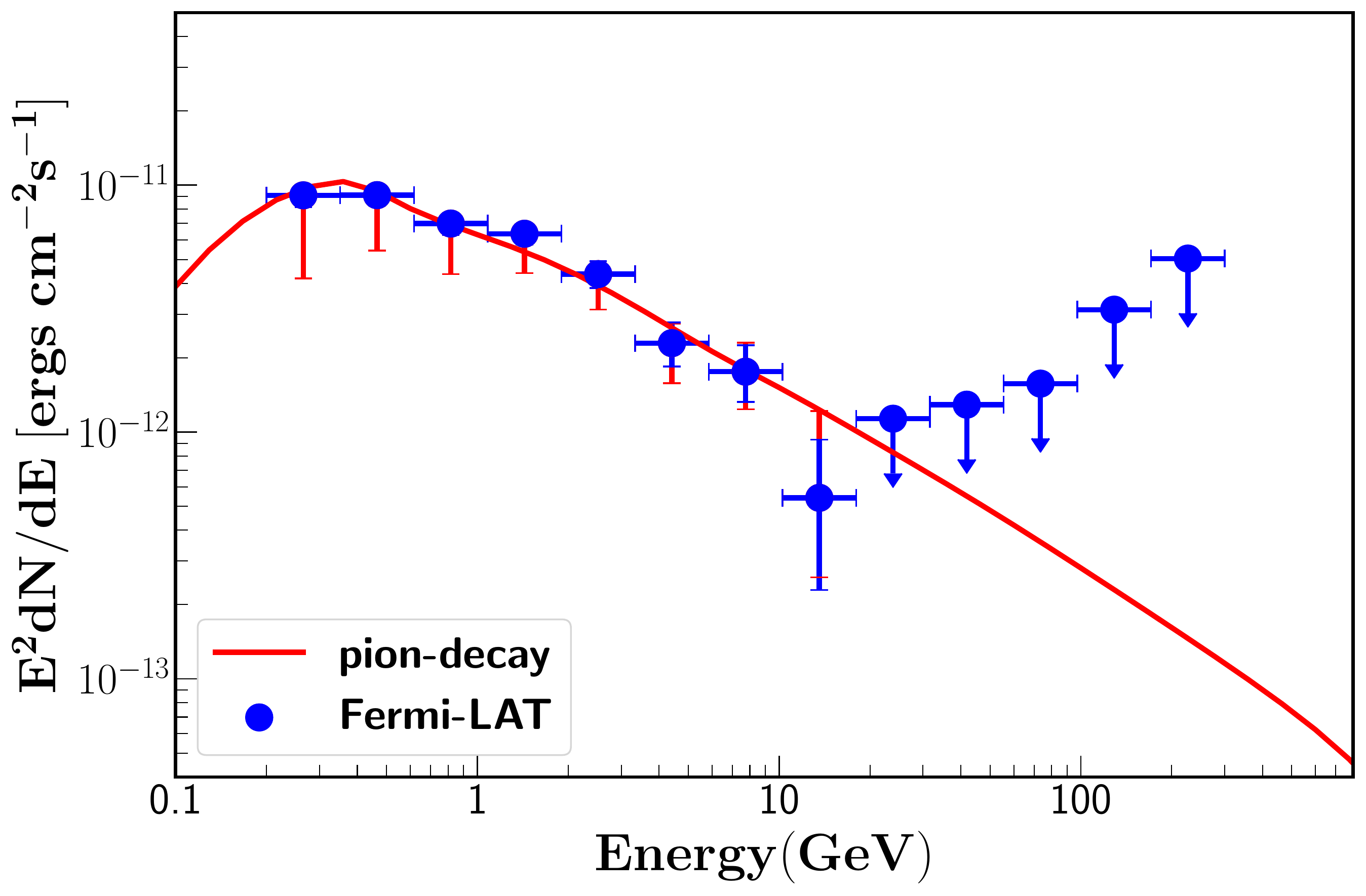}
\end{tabular}
 \caption{The spectral energy distributions and fitted physical models for 3FHL J1907.0+0713/4FGL J1906.9+0712 at 200 MeV $-$ 300 GeV. The parameters of the best-fitting hadronic and leptonic models are given in Table \ref{table_5}. Left: The observed SED is fitted with the leptonic model where both IC and bremsstrahlung emission is first considered to explain the data. The bremsstrahlung emission is calculated for the ambient matter density of 330 cm$^{-3}$. Right: The observed SED is fitted with the hadronic model for the ambient matter density of 330 cm$^{-3}$ (i.e. Hadronic model). The systematic errors are shown with red lines.}
    \label{figure_7}
\end{figure*}

\begin{figure*}
\includegraphics[width=1.0\textwidth]{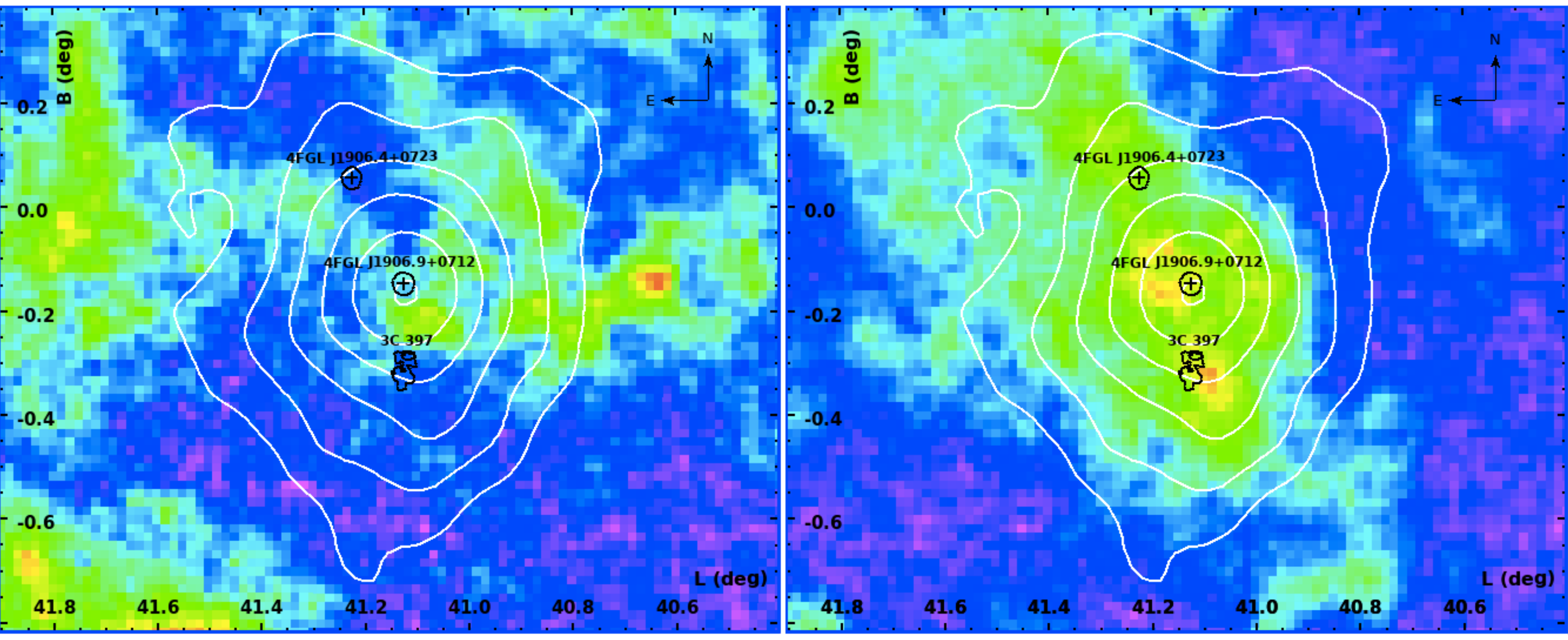}
\caption{The integrated intensity of NANTEN2 $^{12}$CO($J$ = 1$-$0) data with gamma-ray contours produced for the energy range of 1 $-$ 300 GeV. The X-ray data contours of 3C 397 are shown black contours. The positions and their positional errors of PSR J1906+0722/4FGL J1906.4+0723 and 3FHL J1907.0+0713/4FGL J1906.9+0712 are shown with black markers and ellipses. }
    \label{figure_8}
\end{figure*}

\section{Gamma-ray SED modelling of 3FHL J1907.0+0713/4FGL J1906.9+0712} 
\label{Section5}
Gamma-ray emission at MeV $-$ GeV energies is produced through both leptonic and hadronic emission channels. The leptonic emission model includes synchrotron, inverse Compton (IC) and non-thermal bremsstrahlung emission processes, whereas, the gamma-ray emission resulting from the decay of neutral pions (i.e. $\pi^{0}$) is considered as a hadronic emission model. To explain the observed SED of 3FHL J1907.0+0713/4FGL J1906.9+0712, where the SED was produced by applying the best-fit spatial and spectral models (i.e. {\it 2PS Model-2} and Model-1) as discussed in Sections 3.3.1 and 3.3.3, we tested both leptonic and hadronic emission models, which are described in Section \ref{Section5.1} and \ref{Section5.2}, respectively. We use the numerical code developed by \citet{SaBa15} for the physical modelling of the SED. We also use Markov Chain Monte Carlo method \citep{Fo13} to get the best-fit parameters of the physical model.

\subsection{The leptonic model}
\label{Section5.1}
We assume that the observed gamma-ray emission in the range between 200 MeV to 300 GeV results from IC emission and non-thermal bremsstrahlung emission processes. The observed spectrum is best described by a PL model as shown in Table \ref{table_3}. Hence a simple PL model of the energy spectrum of the high energy electrons is presumed which is defined as follows:

$$\rm {dN_e \over dE}= \rm N_0 E^{-\alpha},$$

where $\rm N_e$ denotes the number density of the electrons, E is the energy of the electron, and $\rm N_0$, $\alpha$ are the normalisation, spectral index, respectively. 

The cosmic microwave background (CMB) and Galactic interstellar radiation field are considered as target photons for IC emission. The density of the ambient medium ($n_0$) is accepted to be 330 cm$^{-3}$, which is obtained in Section \ref{Section4.2} and shown in Table \ref{table_5}. 

To obtain the best-fitting spectral parameters we consider both IC and bremsstrahlung process together to account for the observed spectrum. The best-fitting model parameters are shown in Table \ref{table_4}. Fig. \ref{figure_7} left panel shows fit to the data for both IC and bremsstrahlung spectra for the estimated best-fit parameters of the PL electron distribution. It is evident from the left panel of Fig. \ref{figure_7} is that among bremsstrahlung and IC processes, bremsstrahlung is the dominant emission model within the leptonic scenario. The IC contribution is much less than compared to that of bremsstrahlung process. The bremsstrahlung fluxes are proportional to the density of ambient matter. Hence, it is very likely that the bremsstrahlung contribution may dominate over IC process due to high ambient matter density. This is what happened for this specific source.  Following left panel of Fig. \ref{figure_7}, we can conclude that the leptonic model can explain the observed data well for the PL distribution of electrons.

\subsection{The hadronic model}
\label{Section5.2}
In addition to the leptonic model, we also tried to fit the hadronic model to the observed gamma-ray spectrum. Hence we calculated the gamma-ray flux resulting from the decay of neutral pions following the $\pi^0$-decay model of \cite{Kl06} and for a ECPL proton distribution. We used 330 cm$^{-3}$ from Table \ref{table_4} as the ambient matter density. The best-fitting model parameters are given in Table \ref{table_5} and Fig. \ref{figure_7} right panel shows $\pi^0$-decay gamma-ray spectrum for these best-fitting model parameters. It is evident that the hadronic model can also explain the spectral data very well.

\begin{table}
    \caption{\textbf{Parameters of the best-fitting leptonic and hadronic scenarios.}.}
    \begin{minipage}{0.45\textwidth}
    	\centering
	\begin{tabular}{lcccc} 
        \hline\hline
         Model                                             &$\alpha$   &E$\rm_{total}$ & Log-likelihood  \\
                                                                &      &(erg)  &  \\ 
 		\hline
           \\
         \multirow{1}{*}{Leptonic}               & 2.61            & 4.1 $\times$ 10$^{48}$ & -10.2\\
         \\                 
         \multirow{1}{*}{Hadronic}                & 2.80            & 7.8 $\times$ 10$^{46}$ & -3.6\\
        \hline 
        \label{table_5}
        \vspace{-0.3cm}
	\end{tabular}
    \end{minipage}
\end{table}

\section{Discussion} 
\label{Section6}
Fig. \ref{figure_8} shows the off-pulse gamma-ray emission contours for TS values of 25, 49, 81, 121, 169, and 225 on the integrated intensity of NANTEN2 $^{12}$CO($J$ = 1$-$0) data, which are taken in the velocity range of 27.0 $-$ 34.9 km s$^{-1}$ (left panel) and 34.9 $-$ 39.8 km s$^{-1}$ (right panel). 

3FHL J1907.0+0713/4FGL J1906.9+0712 is found to be a point-like gamma-ray source which is in spatial coincidence with GMC B located at about 2.6 kpc. The best-fitting position of the gamma-ray source is at the edge of an arc-shaped MC structure (Fig. \ref{figure_8} right panel), which has an average integrated intensity of approximately 25 K km s$^{-1}$. 

The cosmic proton energy density is calculated to be about 30 eV cm$^{-3}$ by dividing the total proton energy (7.8$\times$10$^{46}$ erg) calculated for the hadronic model given in Table \ref{table_5} into the volume of GMC B at 2.6 kpc, where the diameter of GMC B is given in Table \ref{table_4} as the size. This number is about one order of magnitude larger than the typical number for the ISM, equivalently the CR energy density around the Sun (roughly 1 eV cm$^{-3}$). This may indicate that the CR proton source is nearby or {\it in-situ}. If we have taken the distance of GMC B as 10.2 kpc, we would obtain a very small cosmic proton density, which would be unnatural. Similarly, we calculated the cosmic electron density and found it to be approximately 166 eV cm$^{-3}$ by taking the total electron energy of 4.1$\times$10$^{48}$ erg from Table \ref{table_5} for the leptonic model and the distance to be 2.6 kpc for GMC B. The cosmic electron density seems to be too high to be accepted within the context of the leptonic models.

The observed gamma-ray spectrum is accommodated using both leptonic and hadronic models. We found that the both leptonic and hadronic models can explain the observed SED for ambient matter density of 330 cm$^{-3}$. Within the context of leptonic models, bremsstrahlung emission is the dominant over IC scattering due to the presence of dense MCs. Although both models apparently explain the SED, the log-likelihood value is maximum for the fit of hadronic model to the observed SED as shown in Table \ref{table_5}. The largest improvement with respect to the leptonic model is given by the hadronic model with TS = 13.2, which corresponds to an improvement of only 3.6$\sigma$. This hints that the hadronic model is better compared the leptonic model.

As there is an angular distance of about 0.$^{\!\!\circ}$23 between 3FHL J1907.0+0713/4FGL J1906.9+0712 and PSR J1906+0722/4FGL J1906.4+0723, there is no physical relation between them. However, the location of PSR J1906+0722/4FGL J1906.4+0723 is found to be approximately 1.9 kpc indicating that the pulsar may also be residing inside GMC B or could be related to this MC. On the other hand, the SNR 3C 397 is associated with GMC A, which is at a distance of about 10.7 kpc. So, there is probably no physical connection between 3C 397 and 3FHL J1907.0+0713/4FGL J1906.9+0712, although \citet{Cl15} suggested that the origin of the excess gamma-ray emission may be related to a possible  interaction between the western edge of the SNR and MCs. 

The energy output of 3FHL J1907.0+0713/4FGL J1906.9+0712 is L = 1.1 $\times$ 10$^{34}$ erg s$^{-1}$ at 2.6 kpc distance and L = 1.7 $\times$ 10$^{35}$ erg s$^{-1}$ at 10.2 kpc. Typically, an AGN of type FSQR has a gamma-ray luminosity that ranges from about 10$^{45}$ to 10$^{49}$ erg s$^{-1}$ and average index of about 2.42, while a BL-Lac has a much higher luminosity than an FSQR with a spectral index averaging between 1.9 and 2.2 depending on the sub-class of the BL-Lac object \citep{Aj20}. An AGN with a luminosity of about 10$^{45}$ erg s$^{-1}$ has to be located at approximately 100 Mpc in order to account for the observed luminosity of 3FHL J1907.0+0713/4FGL J1906.9+0712 obtained for the distance of 2.6 kpc. Furthermore, as shown in Fig. \ref{figure_3}, we detected no significant variability in 3FHL J1907.0+0713/4FGL J1906.9+0712. As a result, the possibility of 3FHL J1907.0+0713/4FGL J1906.9+0712 having an extragalactic origin is ruled out. 

In addition, although the position of the H\,{\sc ii}  region (G41.1-0.2) mentioned in Section \ref{Section1} is close to 3FHL J1907.0+0713/4FGL J1906.9+0712, they are not physically related to each other, because in contrast to 3FHL J1907.0+0713/4FGL J1906.9+0712 being at about 2.6 kpc, the H\,{\sc ii}  region is located at the far-side at about 10.5 kpc. The second closest H\,{\sc ii} region (G41.23-0.19) seems to be spatially coinciding with GMC B, but the derived distances of this SFR shows that this region may physically not be related to 3FHL J1907.0+0713/4FGL J1906.9+0712.

We also searched for a multi-wavelength counterpart of 3FHL J1907.0+0713/4FGL J1906.9+0712 within a circular region of 2 arc-minutes radius in the SIMBAD database\footnote{http://simbad.u-strasbg.fr/simbad/sim-fcoo}. The closest object found to 3FHL J1907.0+0713/4FGL J1906.9+0712 was a star named 2MASS J19070397+0711398 \citep{Cu03}, which is only 5.4 arc-seconds away [($\alpha$J2000, $\beta$J2000) = (19h 07m 03.97s, 07$^{\circ}$ 11$^{\prime}$ 39.86$^{\prime\prime}$)] from the best-fitting location of the gamma-ray source. This star is the central star for the bow-shock nebula candidate, called G041.1120-00.1753, reported by \citet{Ko16}.

3FHL J1907.0+0713/4FGL J1906.9+0712 is possibly a Galactic gamma-ray source by looking at its location and luminosity. Since this source is spatially coincident with GMC B that has a density of 330 cm$^{-3}$, it could be located within a star-forming region considering that the pulsar and the bow-shock nebula are in the close vicinity. 

In such a scenario, the candidate source class could be an SNR, which has not been detected in radio continuum or other wavelengths yet. However, the possibility of a pulsar wind nebula (PWN) to be the source of gamma-ray emission detected from 3FHL J1907.0+0713/4FGL J1906.9+0712 is low, because there is no spatial overlap between a pulsar and this source. In addition, there are no TeV gamma-ray and/or X-ray and radio nebula that can spatially be associated with 3FHL J1907.0+0713/4FGL J1906.9+0712 for supporting a PWN origin of this source.

Other alternatives are: (1) Massive protostars associated with bipolar outflows \citep{Ar08,Bo10} producing strong shocks when they interact with the ambient medium; (2) systems of colliding stellar winds \citep{Re06,Ps16}; (3) bow-shocks of runaway stars \citep{Va13, Sc14}; (4) young star clusters.
 
Dense molecular gas hosts regions of on-going star formation, where gamma rays may be formed from particles accelerated to multi-TeV energies at the shocks produced by outflows from massive protostars. The kinetic luminosity of one such jet is expected to be around 10$^{36}$ erg s$^{-1}$. Both the reverse shock and the bow shock may contribute to the non-thermal radiation and the bow-shock luminosity may be roughly 10$^{34}$ erg s$^{-1}$ \citep{Bo10}. The luminosity of 3FHL J1907.0+0713/4FGL J1906.9+0712 found in this study (1.1$\times$10$^{34}$ erg s$^{-1}$ at 2.6 kpc) is comparable to this value. It is estimated that at high density MCs, relativistic bremsstrahlung and proton-proton collisions can dominate.

The collision of supersonic winds of early-type stars (O, early B, Wolf-Rayet [WR] stars) are hot stars (T $>$ 10000 K$^{\circ}$) having masses of 20 solar masses or higher and they are expected to produce strong shocks, where both electrons and protons can be efficiently accelerated to high energies through first-order Fermi acceleration \citep{Re06}. In the case of WR stars, the wind kinetic power of these stars might be higher than roughly 10$^{34}$ erg s$^{-1}$, where some fraction of this energy is injected into the wind-wind interaction of the system \citep{Ps16}. So far, $\eta$ Car \citep{Ta09, Be11, Re15} and W11 \citep{Ps16} were the only two colliding wind binary systems, from which high energy gamma rays were detected. Although $\eta$ Car system detected by AGILE showed a two-day gamma-ray flaring episode, W11 presented no variability. The average luminosity of the $\eta$ Car system was found to be 3.4 $\times$ 10$^{34}$ erg s$^{-1}$ at 2.3 kpc and for W11 system it was calculated to be 3.7 $\times$ 10$^{31}$ erg s$^{-1}$ at 340 pc. 

Strong winds of runaway OB stars sweep relatively large amounts of interstellar material forming bow shocks, which are observed as arc-shaped features in front of the stars, while they move supersonically in the surrounding ISM. \citet{Be10} suggested that bow shocks are emitters of high-energy gamma rays (E $>$ 100 MeV). 

Young star clusters are reported as candidates for gamma-ray emission, because they supply sufficient kinetic energy through colliding stellar winds for the interactions of cosmic rays with surrounding gas \citep{Ac11, Ab12, Ah19, Sa20}. Some of these star-forming regions are more extended in size and are associated with diffuse gamma-ray emission, which is not the case for 3FHL 1907.0+713/4FGL J1906.9+0712. We found 3FHL 1907.0+713/4FGL J1906.9+0712 as a point-like source that is similar to the point-like nature of the 4FGL J1115-6118 which is associated with the star-forming region NGC 3603 \citep{Sa20}. The hadronic origin of gamma rays in of both of these sources is a favourable emission mechanism due to the presence of dense MCs. This also hints that star formation could be the origin of gamma rays for this source. 

\section{Conclusions}
\label{Section7}
In this paper, we studied the nature of 3FHL J1907.0+0713/4FGL J1906.9+0712 and the relation to its environment and neighbouring sources. Below are the main outcomes of this study: 

\begin{itemize}
    \item In the energy range of 0.2 $-$ 300 GeV, we detected 3FHL J1907.0+0713/4FGL J1906.9+0712 as a point source and with a significance of about 13$\sigma$ after the removal of gamma-ray pulsation periods of PSR J1906+0722/4FGL J1906.4+0723 from the {\it Fermi}-LAT data set of about 10 years. It's best-fitting location was found to be (RA, Decl.) =  (286$^{\circ}\!\!$.77 $\pm$ 0$^{\circ}\!\!$.02, 7$^{\circ}\!\!$.19 $\pm$ 0$^{\circ}\!\!$.01).
        
    \item We found no significant variability in the 1-month-binned gamma-ray light curve of 3FHL J1907.0+0713/4FGL J1906.9+0712 and within the circular region of 0.$\!\!^{\circ}$1 radius surrounding this source in the energy range of 200 MeV and 300 GeV. 
       
    \item In the energy range of 0.2 $-$ 300 GeV, the spectrum of 3FHL J1907.0+0713/4FGL J1906.9+0712 is well-fit with a  PL-type spectrum with an index of $\Gamma$ = 2.26 $\pm$ 0.05. The total photon flux was calculated to be (1.03 $\pm$ 0.11)$\times$10$^{-8}$ cm$^{-2}$ s$^{-1}$ and the total energy flux value was found to be (8.51 $\pm$ 0.78) $\times$ 10$^{-6}$ MeV cm$^{-2}$ s$^{-1}$. 
 
     \item The off-pulse emission of PSR J1906+0722/4FGL J1906.4+0723  was detected in the energy range of 0.2 $-$ 300 GeV at a significance level of about 14$\sigma$. Its spectrum was best-fitting to a PL-type spectrum having a spectral index of 2.65 $\pm$ 0.06. The photon and energy fluxes were found to be (1.74 $\pm$ 0.14) $\times$ 10$^{-8}$ cm$^{-2}$ s$^{-1}$ and (8.71 $\pm$ 0.66) $\times$ 10$^{-6}$ MeV cm$^{-2}$ s$^{-1}$. However, in the energy range from 1 to 300 GeV of the off-pulse emission, PSR J1906+0722/4FGL J1906.4+0723 could not be detected. 
 
    \item No significant gamma-ray emission was detected from the SNR 3C 397 in the 0.2 $-$ 300 GeV energy range of the off-pulse data.
    
    \item 3FHL J1907.0+0713/4FGL J1906.9+0712 was found to be spatially coincident with the MC called GMC B, which has a peak velocity of about 38 km s$^{-1}$ and an estimated distance of approximately 2.6 kpc. The density at 2.6 kpc was calculated to be 330 cm$^{-3}$. The best-fitting position of the gamma-ray source is at the edge of an arc-shaped structure of GMC B. The total luminosity of 3FHL J1907.0+0713/4FGL J1906.9+0712 at a distance of 2.6 kpc was calculated to be 1.1 $\times$ 10$^{34}$ ergs s$^{-1}$. 

    \item The gamma-ray SED modelling of 3FHL J1907.0+0713/4FGL J1906.9+0712 showed that although both hadronic and leptonic gamma-ray emission scenarios can explain the current gamma-ray emission from this source, the log-likelihood fit done for the hadronic model resulted in an improvement of about 3.6$\sigma$ over the fit result obtained for the leptonic model.
    
    \item Although the angular separation between PSR J1906+0722/4FGL J1906.4+0723 and 3FHL J1907.0+0713/4FGL J1906.9+0712 is found to be 0$^{\circ}\!\!$.23, the distance of the pulsar (roughly 1.9 kpc) may indicate that, like 3FHL J1907.0+0713/4FGL J1906.9+0712, it may be related to GMC B. On the other hand, although the best-fitting location of 3FHL J1907.0+0713/4FGL J1906.9+0712 is 0$^{\circ}\!\!$.13 away from 3C 397's position, 3C 397 was found to reside inside another MC called GMC A, the distance of which was estimated to be approximately 10.7 kpc. Therefore, these three sources are probably physically not related to each other.   
                  
    \item The closest object found to 3FHL J1907.0+0713/4FGL J1906.9+0712 was a star called 2MASS J19070397+0711398, which is only 5.4 arc-seconds away from the gamma-ray source. This star is the central star associated with a bow-shock nebulae. Because dense MCs are thought to be places of star formation, we concluded that possible candidate source classes for 3FHL J1907.0+0713/4FGL J1906.9+0712 may be (1) yet-undetected SNRs interacting with MCs, (2) massive protostars associated with bipolar outflows, (3) strong winds of runaway OB stars, (4) systems of colliding stellar winds, and finally (5) young star clusters. For the latter 4 cases, the strong stellar winds and their interactions with the surrounding medium or with another star's wind may play a crucial role in the production of high energy gamma rays.         
\end{itemize}

\vspace{-0.3cm}
\section*{Acknowledgements}
The authors are thankful to Dr. C. J. Clark, Albert-Einstein-Institut, MPIG, Germany and to Dr. Matthew Kerr, Space Science Division, Naval Research Laboratory, Washington, USA for providing the Ephemeris of PSR J1906+0722. We are particularly grateful to Dr. Matthew Kerr also for providing the guidance of generating the pulsar phase plot in python. L.S. acknowledges financial support of the ERDF under the Spanish MINECO (FPA2015-68378-P and FPA2017-82729-C6-3-R). P.B. is thankful to the DST-INSPIRE Fellowship scheme. R.Y. and S.J.T. deeply appreciate Aoyama Gakuin University Research Institute for helping our research by the fund. This work was supported by JSPS KAKENHI Grant Numbers JP17H18270 (S.J.T.), JP18H01232 (R.Y.), JP19K14758 (H.S.), and JP19H05075 (H.S.). The NANTEN project is based on a reciprocal agreement between Nagoya University and the Carnegie Institution of Washington. We deeply acknowledge that the NANTEN project was realised by contributions from many Japanese public donors and companies. 

\vspace{-0.3cm}
\section*{Data Availability}
The {\it Fermi}-LAT data underlying this article are available at \url{https://fermi.gsfc.nasa.gov/ssc/data/access/lat/}. The $^{12}$CO($J$ = 1$-$0) data used in this study will be made available by the corresponding authors upon request.



\label{lastpage}
\end{document}